\documentclass[12pt]{iopart}

\usepackage{hyperref}
\usepackage{xspace}
\usepackage{graphicx}
\usepackage{cite}
\usepackage{subcaption}
\usepackage{stfloats}
\usepackage{smartdiagram}
\usepackage{soul}
\usepackage[normalem]{ulem}
\usepackage{xcolor}
\usepackage{savesym}
\savesymbol{comment}  
\usepackage{changes}

\usepackage[nolist,nohyperlinks]{acronym}
\acrodef{BIM}[BIM]{\emph{Building Information Modeling}}
\acrodef{CSR}[CSR]{\emph{Cholesteric Spherical Reflector}}
\acrodef{CLC}[CLC]{\emph{Cholesteric Liquid Crystal}}
\acrodef{AR}[AR]{\emph{Augmented Reality}}
\acrodef{AEC}[AEC]{\emph{Architecture, Engineering and Construction}}
\acrodef{SKU}[SKU]{\emph{Stock Keeping Unit}}
\acrodef{CMYK}[CMYK]{\emph{Cyan Magenta Yellow Black}}
\acrodef{NFC}[NFC]{\emph{Near Field Communication}}
\acrodef{CBC}[CBC]{\emph{Construction Blockchain Consortium} } 
\acrodef{IFC}[IFC]{\emph{Industry Foundation Classes}}
\acrodef{SLAM}[SLAM]{\emph{Simultaneous Localization And Mapping}}
\acrodef{SKU}[SKU]{\emph{Stock Keeping Unity}}
\acrodef{PLM}[PLM]{\emph{Product Lifecycle Management}}
\acrodef{PUF}[PUF]{\emph{Physical Unclonable Function}}
\acrodef{V2V}[V2V]{\emph{Vehicle To Vehicle}}
\acrodef{V2I}[V2I]{\emph{Vehicle To Infrastructure}}
\acrodef{I2V}[I2V]{\emph{Infrastructure To Vehicle}}
\acrodef{AI}[AI]{\emph{Artificial Intelligence}}
\acrodef{IOT}[IoT]{\emph{Internet of Things}}
\acrodef{NOA}[NOA]{\emph{Norland Optical Adhesive}}

\acrodef{UV}[UV]{Ultraviolet\xspace}
\acrodef{IR}[IR]{Infrared \xspace}

\newcommand{\eg}{\emph{e.g.},\xspace}
\newcommand{\ie}{\emph{i.e.},\xspace}

\begin{document}

\title[Linking Physical Objects to Their Digital Twins with Invisible Fiducial Markers]{Linking Physical Objects to Their Digital Twins via Fiducial Markers Designed for Invisibility to Humans}

\author{Mathew~Schwartz$^{1*}$, Yong Geng$^{2\dagger}$, Hakam Agha$^{2\dagger}$, Rijeesh Kizhakidathazhath$^2$, Danqing Liu$^3$, Gabriele Lenzini$^4$ and Jan~PF~Lagerwall$^{2*}$}

\address{$^1$New Jersey Institute of Technology, College of Architecture and Design,\\University Heights, Newark, NJ, USA\\
$^2$University of Luxembourg, Department of Physics and Materials Science,\\162a Avenue de la faiencerie, L-1511 Luxembourg city, Luxembourg\\
$^4$Eindhoven University of Technology, Department of Chemical Engineering and Chemistry, Den Dolech 2, 5612 AZ, Eindhoven, The Netherlands\\
$^3$University of Luxembourg, Interdisciplinary Centre for Security and Trust, L-4365 Esch sur Alzette, Luxembourg\\
}
\eads{\mailto{cadop@umich.edu},\mailto{jan.lagerwall@lcsoftmatter.com}}
\vspace{10pt}
\begin{indented}
\item{$^{\dagger}$These authors contributed equally.}
\item[]January 2021
\end{indented}

\vspace{2pc}

\begin{abstract}
The ability to label and track physical objects that are assets in digital representations of the world is foundational to many complex systems. Simple, yet powerful methods such as bar- and QR-codes have been highly successful, \eg in the retail space, but the lack of security, limited information content and impossibility of seamless integration with the environment have prevented a large-scale linking of physical objects to their digital twins. This paper proposes to link digital assets created through \ac{BIM} with their physical counterparts using fiducial markers with patterns defined by \textit{Cholesteric Spherical Reflectors} (CSRs), selective retroreflectors produced using liquid crystal self-assembly. The markers leverage the ability of CSRs to encode information that is easily detected and read with computer vision while remaining practically invisible to the human eye. We analyze the potential of a CSR-based infrastructure from the perspective of \ac{BIM}, critically reviewing the outstanding challenges in applying this new class of functional materials, and we discuss extended opportunities arising in assisting autonomous mobile robots to reliably navigate human-populated environments, as well as in augmented reality. 

\end{abstract}

\section{Introduction}\label{preamble}
In our contribution to this special issue on \textit{Morphological Computing}, we have chosen to take a viewpoint that is complementary to the focus on functional materials for robots, discussing a novel functional material integrated in the \textit{environment} with which robots---and humans---interact. We argue that the impact of this material in enabling autonomous mobile robots and related emerging technology to transform our modern world, by helping them to better make sense of the environment, can be profound. This resonates well with the definition of morphological computation used by Pfeifer \cite{Pfeifer2006}, in which the connection to the environment is a central component.

\begin{figure}[b!]
\centering
\includegraphics[page=2, width=\textwidth]{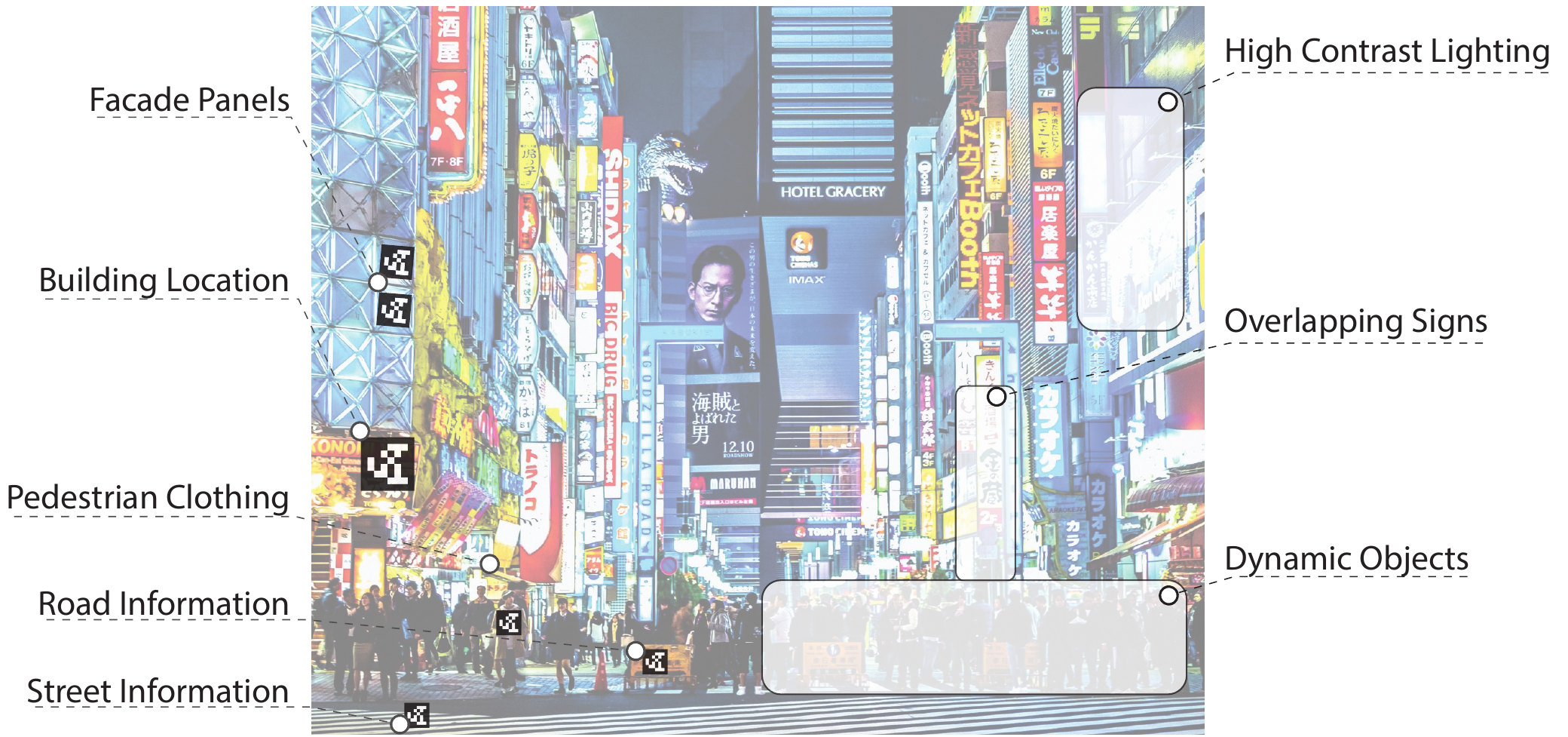} 

 \caption{An example of a modern cityscape (Tokyo, Japan), with the right side highlighting elements dense in visual information and posing numerous problems for localization and understanding the environment. The left side of the figure illustrates where and how fiducial markers would be beneficial. Modified photo from\cite{unsplash-tokyo} }
\label{fig:citycenter}
\end{figure}

Today's human-populated environments are largely off-limits for robots due to safety concerns, very much related to the technological challenges associated with making robots understand their surroundings \cite{Li2020,Zhang2020b,Klomp2019,Hamid2018}. There is good reason to address these challenges, because a future of a fully sensored and augmented environment to which our devices, including robots co-occupying our spaces, are linked in their knowledge of location, place, and identity, could provide many benefits to society, ranging from enhanced personal mobility to low ecological footprint technologies \cite{Kim2016d,Iglinski2017,Alfeo2019,Jones2020}. To realize this vision, the ability of robots to correctly interpret existing complex and dynamic cityscapes (figure~\ref{fig:citycenter}) would either need to be dramatically improved, or the environment should be adapted to the robots. The former approach entails significantly enhanced collection and processing of data from environments with human occupants, raising personal integrity concerns related to, \eg face recognition \cite{Castelvecchi2020}. It would also come at the cost of increased on-board computation or heavy reliance on potentially insecure wireless communication and competition for radio frequency bandwidth.

In this paper we advocate the second approach, \ie a redesign of the environment by adding artificial landmarks to indoor and outdoor spaces, as well as to the objects they contain. These landmarks should provide reliable, secured and easily readable information to robots, allowing them to understand where they are, where to go, and to determine which other actors are in their surroundings. The idea is not to replace existing navigation technologies, but to support them by adding a powerful set of tailored signals, enhancing the reliability and correctness of robotic navigation and reducing computational needs. Importantly, our proposed solution allows this redesign to be without consequences for the environment as perceived by humans, since the landmarks should be invisible to the human eye.  

\begin{figure}[t!]
\includegraphics[page = 7, width=16cm]{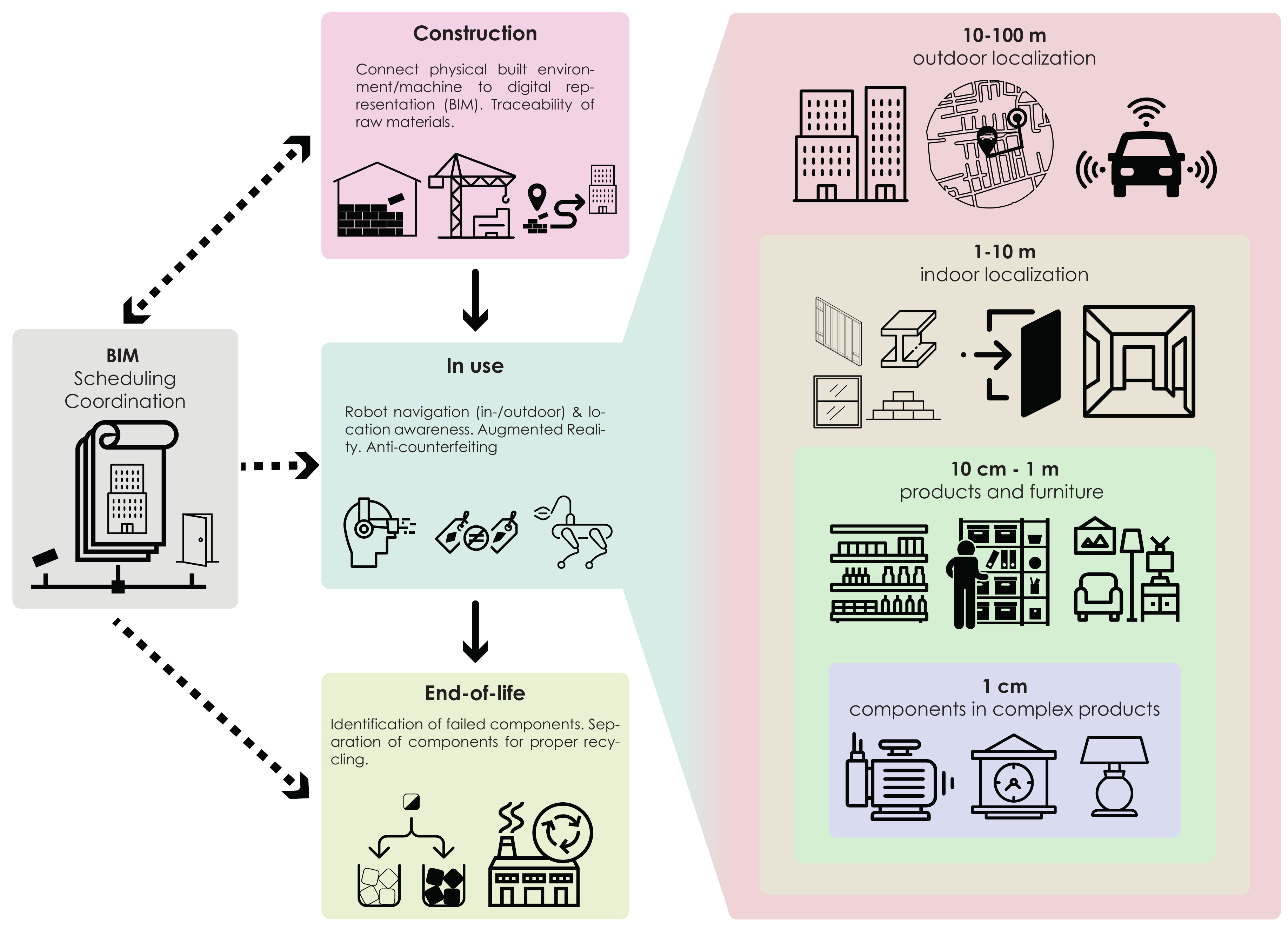} 
\caption{An overview of the opportunities opened, on multiple scales, by ubiquitous links between physical objects and their digital twins realized by CSR-enabled unobtrusive fiducial markers. Each scale covered is coordinated through BIM, throughout each object life cycle.}
\label{fig:bim_scales}
\end{figure}

The artificial landmarks that we consider create links between the physical objects carrying the landmarks and their digital representations. Such links are extremely valuable well beyond the realm of robotic navigation. Multiple disciplines are currently exploring ground-breaking technologies to interconnect the physical and the digital worlds, a transformation that our links could help realize, with revolutionary benefits across a large spectrum of use cases (see figure~\ref{fig:bim_scales} and section~\ref{digrepresent}). By giving every physical element in the built environment a digital identity, an approach standardized within \textit{Building Information Modeling}, or \ac{BIM}~\cite{eastman2011bim}, even complex physical spaces can be represented and managed digitally, with high resolution in the data pipeline. Each identity in the computer representation can then be referred to as the \textit{Digital Twin} of the corresponding physical element, a term introduced within the scope of \ac{PLM}~\cite{grieves2005product}. Recent developments focused on software-side breakthroughs such as blockchain to keep records and track building components throughout the life cycle, from acquisition to installation to recycling. However, a convenient link between the physical component and its digital twin---before, during and after construction---is still missing. Ideally, this link would be provided unobtrusively and passively, minimizing the need for electronic beacons or object-embedded \ac{IOT} devices that rely on electrical power. While \ac{NFC} technology has been suggested for interaction with drone swarms \cite{Jones2020}, the requirement for physical proximity during read-out limits the scaleability and general usefulness.

\begin{figure}
\centering
\includegraphics[page=3, width=\textwidth]{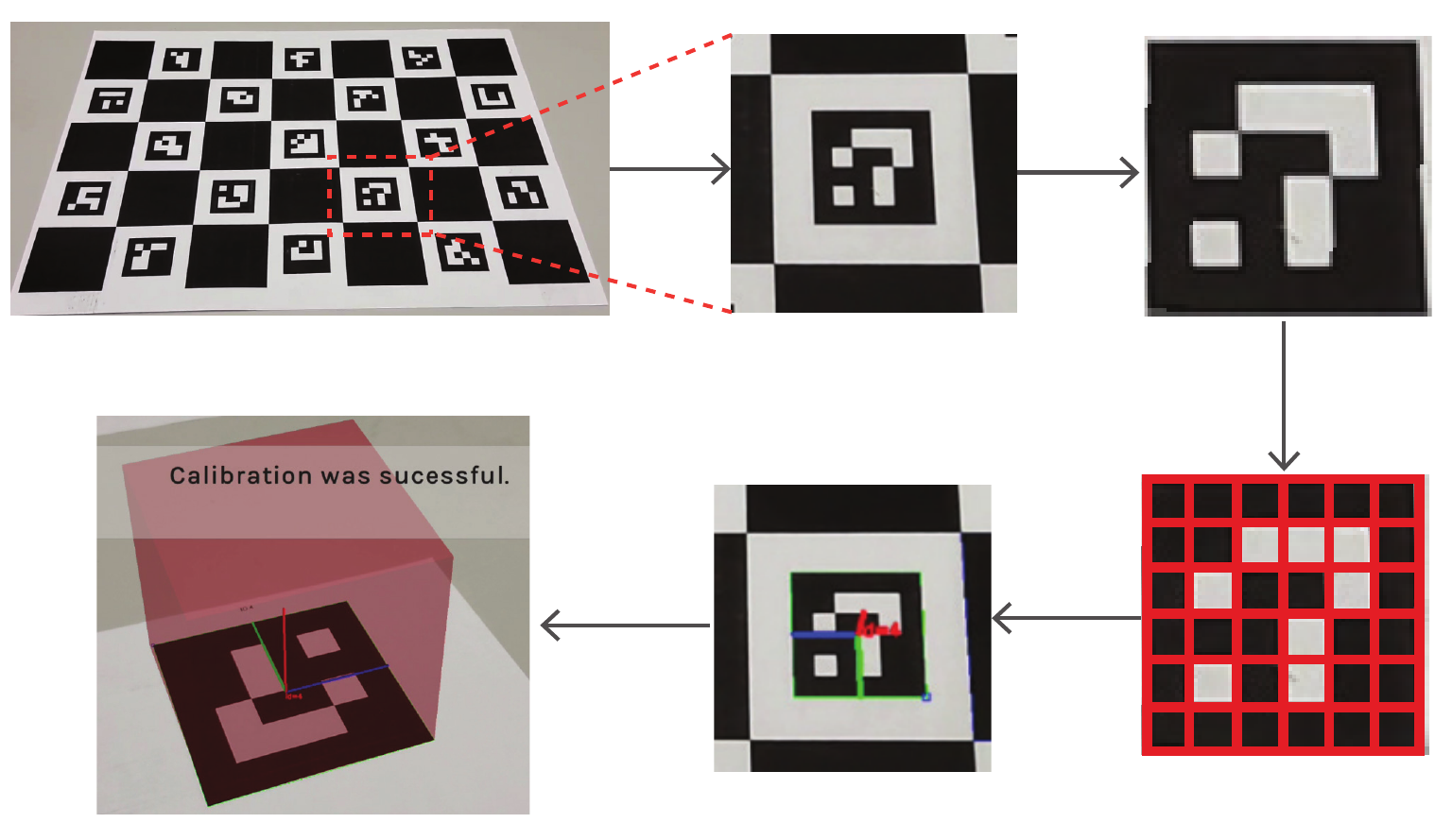} 
\caption{Illustration of how fudicial markers are read and understood with computer vision. After taking an image and identifying markers in the image, the markers are undistorted digitally. Then, the grid that the original marker was defined on is recreated, with each cell in the grid described as a binary value. Finally, the 3-dimensional axis set is projected on the marker (homography), which can be used for localization and understanding in 3D space. The last image (bottom left) shows the marker in a new orientation, while the projected axis remains and a virtual 3D cube is projected into the real-world space (\ac{AR}). }
\label{fig:markers_overview}
\end{figure}

In closed-off technical environments (research labs, high-risk facilities like nuclear power plants, military facilities, \dots), the link is provided very efficiently using a class of artificial landmarks referred to as \textit{fiducial markers}~\cite{Garrido-Jurado2014,zhang2015localization,Sagitov2017,Krogius2019}. These are binary rectangular patterns of known size, typically consisting of black squares on a white background, see figure~\ref{fig:markers_overview}. QR-codes can be compared to fiducial markers, although their patterns are more complex and richer in information. Because a robot can establish the distance and orientation to a fiducial marker from a simple calibrated 2D camera image, by analyzing the overall size and shape change as it depends on the distance and perspective of readout (figure~\ref{fig:markers_overview}), the deployment of fiducial markers in the surroundings ensures successful localization. Transferring the fiducial marker-based solution to human-populated spaces, and thereby realizing the ubiquitous links between the physical environment and its digital representation, would be highly desirable, but the markers would then need to be unobtrusive to the environment as experienced by humans, since markers would be everywhere. This rules out traditional fiducial markers, highly visible and functioning only under sufficient illumination. Moreover, they may be less efficient in visually complex environments like cityscapes, where there are many other patterns that could incorrectly be interpreted as fiducial markers, giving false positives.

We recently presented a new material component that we call \ac{CSR} \cite{schwartz2018cholesteric,Geng2021encoding} that would allow the realization of fiducial markers that robots can easily detect and analyze, with no impact on the environment as we see it. CSRs exhibit selective omnidirectional retroreflectivity, with distinct right- or left-handed circular polarization, in a narrow wavelength band that can be tuned to the invisible \ac{UV} or \ac{IR} spectra to ensure that they are undetectable by the human eye. This means that fiducial markers made using CSRs (in the following referred to as CSR fiducials, for simplicity) would provide a solution that combines the direction independence of motion capture technology (normally based on \textit{visible} retroreflectors) with the hidden messages of the world of flowers, where rich patterns shining vividly in the \ac{UV} and \ac{IR} spectra guide bees and other pollinating insects, yet these signals remain oblivious to us \cite{Koski2014,Klomberg2019}. Note that the UV range considered is the very near-UV range that is harmless to humans and is already used (under the name ``blacklight'') in many contexts, like night clubs or in watermark visualization. The circular polarization of the reflections allows exceptionally simple image segmentation, \ie removal of the background, thereby removing the risk of false positives \cite{Geng2021encoding}.

We focus particularly on the way in which CSR fiducials can support the implementation of \ac{BIM} in construction of the built environment, by providing the links between the physical and digital worlds. With CSR fiducials being incorporated in the environment as it emerges, the construction industry would immediately benefit and an infrastructure would gradually develop that can eventually assist robots of all kinds. This strategy largely avoids the enormous logistical challenge of marking up existing environments.

\section{The Working Principle of CSRs and their Implementation in Fiducial Markers} \label{csr:sec}

\begin{figure}[b!]
    \begin{center}
    \includegraphics[width=9cm]{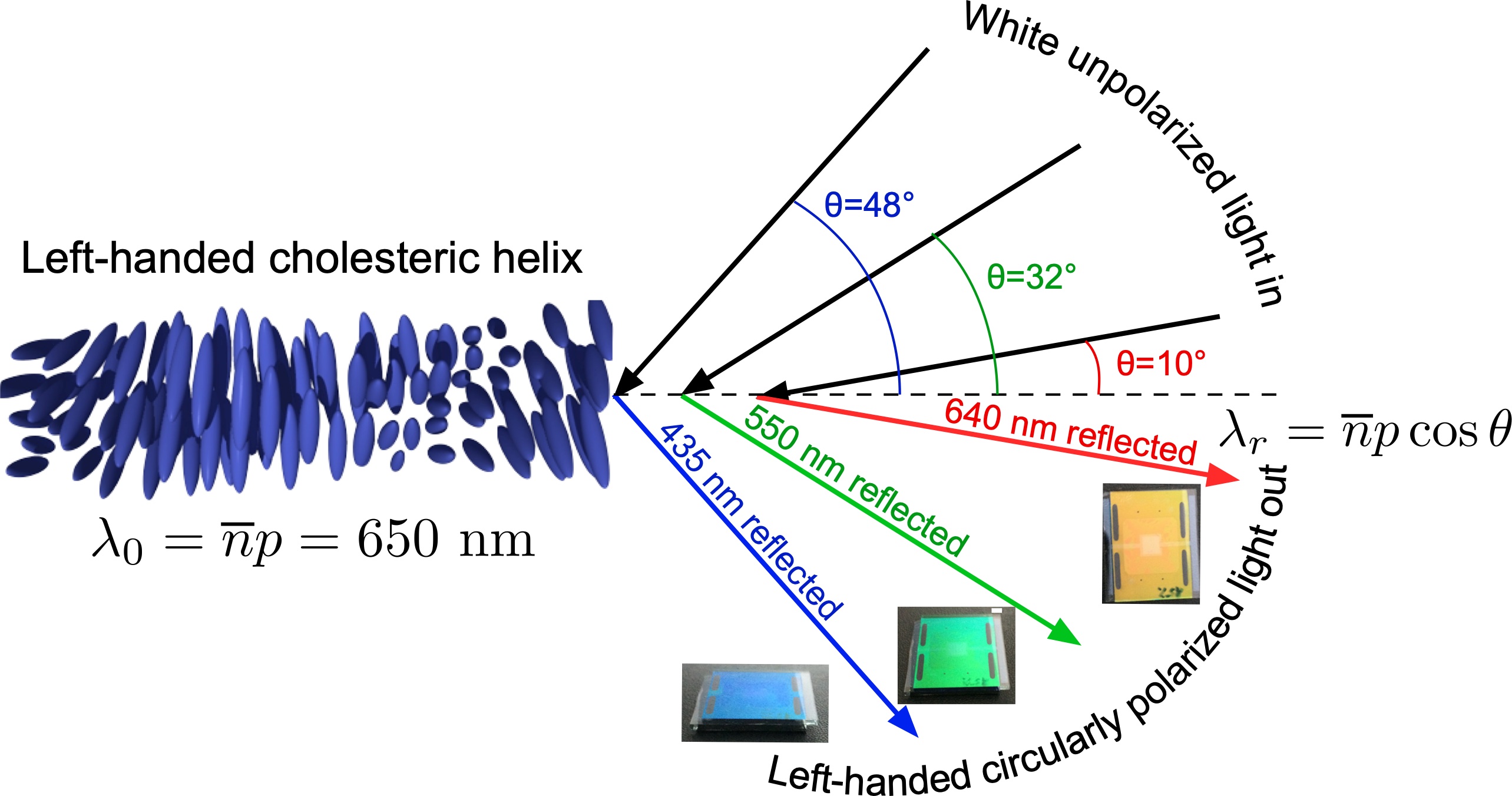} 
     \caption{The cholesteric helix gives rise to circularly polarized and angle-dependent wavelength-selective (thus colored) Bragg reflection. This example is for a pitch $p$ yielding orthogonal reflection (retroreflection) $\lambda_0$ in the visible red.}
    \label{fig:helix}
    \end{center}
\end{figure} 
CSRs are made by bringing cholesteric liquid crystals (CLCs) into spherical shape, in form of a droplet \cite{Humar2010,noh2014tuneable, Asshoff2015, Fan2015, lee2017structural,Wang2017a} or a shell \cite{uchida2013controlled,geng2016high,lee2017structural,geng2017elucidating,schwartz2018cholesteric,myung2019optical,iwai2020shrinkage,Park2020}. The latter is a droplet of CLC which itself contains a droplet of immiscible isotropic liquid. In contrast to ordinary liquids, CLCs are anisotropic; their molecules self-organize with long-range orientational order, modulated into a helical structure with a period (pitch, $p$) of $\sim$0.1~$\mu$m or longer \cite{Bisoyi2018}. This turns CLCs into fluid Bragg reflectors, with selective reflection in a narrow wavelength band that blue-shifts with increasing viewing angle $\theta$ with respect to the helix axis \cite{Kitzerow2000}, see figure~\ref{fig:helix}. The reflection wavelength $\lambda_r$ depends also on $p$ and the average refractive index $\bar{n}$ in the CLC, following Bragg's law: $\lambda_r=\bar{n}p\cos{\theta}$. Importantly, the Bragg-reflected light is circularly polarized, with the same handedness as the CLC helix. Because we can choose helix handedness and adjust $p$ by varying the composition of a CLC mixture, we can select left- or right-handed polarization and tune the band of \textit{retro}reflection---corresponding to $\theta = 0$ and thus the longest reflection wavelength $\lambda_0=\bar{n}p$---from infrared (IR), through the visible spectrum, to ultraviolet (UV).

For flat CLC samples, a problem is the viewing angle dependence of $\lambda_r$: people observing a flat CLC along different angles will not agree on its color. A robot scanning for a certain CLC reflection may miss the signal for the same reason. However, with a spherical CLC with radial helix, \ie a CSR, you are looking along the CLC helix at the center of the CSR no matter how you observe it. Illuminating along the viewing direction, you will thus always have selective retroreflection at $\lambda_0$ \cite{Humar2010,uchida2013controlled,schwartz2018cholesteric}. This means that a fiducial marker where one set of pixels (typically those that normally are black) are made of CSRs reflects the designed pattern back to an observer, irrespective of viewing angle. It thus shares the omnidirectional retroreflectivity of high visibility (HV) clothing, motion capture markers and traffic signage, but in strong contrast to the indiscriminate glass beads or prisms in standard retroreflectors, a CSR sends the light back only if it has the correct wavelength and polarization.

\begin{figure}[t!]
   \centering
   \includegraphics[width=.7\textwidth]{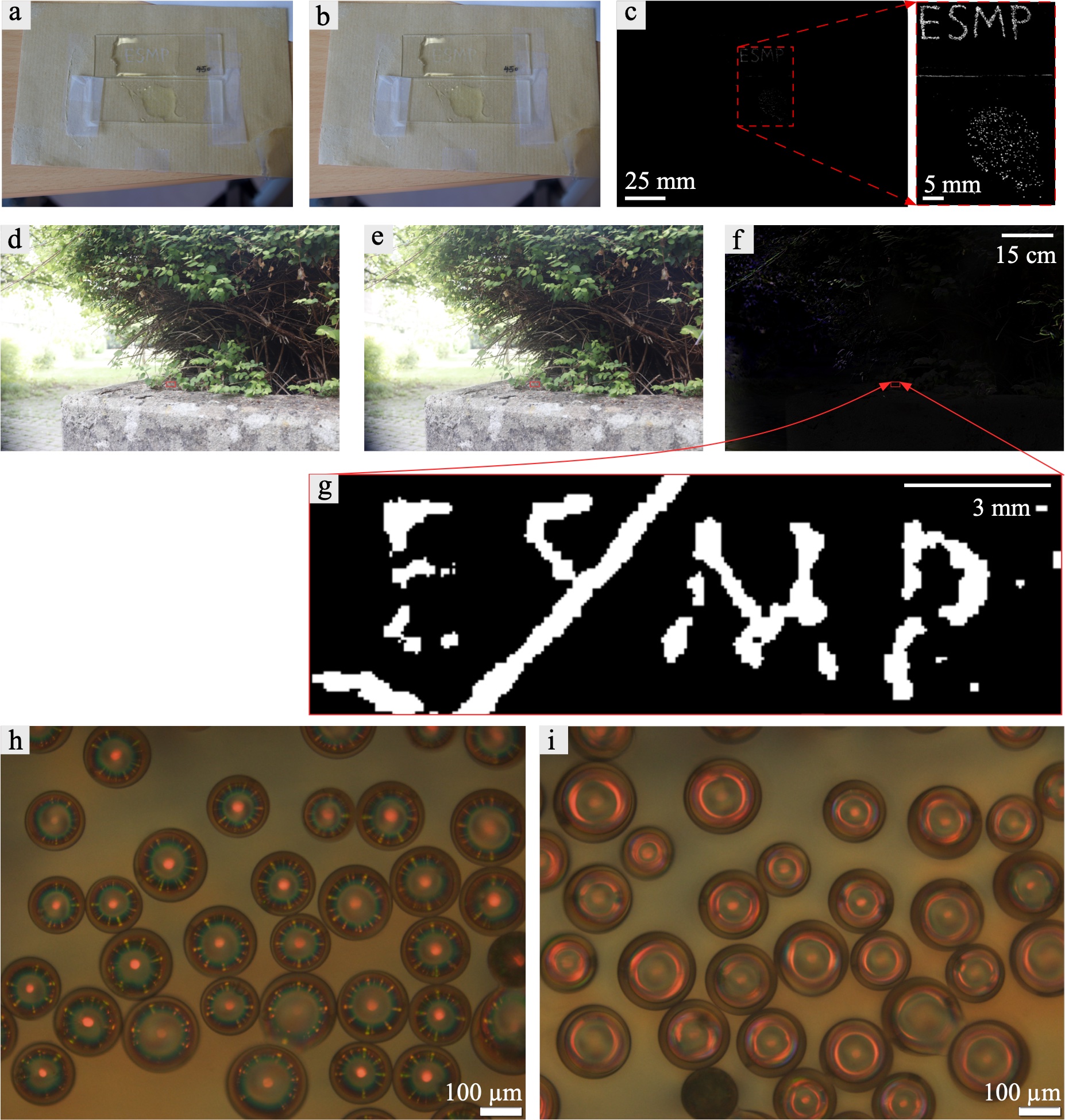}
   \caption{(a--g) Subtraction of background to reveal information encoded into coatings by CSRs in two different environments. Two coatings with blue-reflecting CSRs (a-c; top: tracing the letters ESMP; bottom: random arrangement) are photographed through right- (a) and left-handed (b) circular polarizers. The latter is subtracted from the former and the blue channel extracted in (c), followed by contrast enhancement and magnification (right). The same procedure is shown in (d) to (g) for deep red-reflecting CSRs in an outdoor scene, photographed at several meters distance. The slight movement of a thin branch between the two photos prevents its successful removal, leaving the branch visible between the 'S' and 'M'. The near-field response of asymmetric shell CSRs with $\lambda_0\approx900$~nm are shown in the bottom row, imaged through a polarizing optical microscope from the thick side in (h) and from the thin side in (i). The central retroreflection spot is reddish despite the large value of $\lambda_0$ due to the large opening angles of the illumination and imaging cones, yielding reflection at $\theta>0$.}
   \protect{\label{fig:clc_scales}}
\end{figure}

The circular polarization of the reflection is a powerful characteristic, because it allows us to identify a \ac{CSR} fiducial even if it is hidden in a visually busy environment. In figure~\ref{fig:clc_scales}, three coatings containing CSRs are placed on different backgrounds and photographed using a standard Canon EOS 77D digital SLR camera. First (panels a/d), the photo is taken through a right-handed circular polarizer (we used the right eye of a pair of standard 3D cinema goggles), transmitting the reflection of these right-handed CSRs. Then, without moving the camera, we take the corresponding photo through a left-handed circular polarizer (the left eye of the pair of 3D cinema goggles), which blocks the CSR reflection (panels b/e). Some scattering due to coating imperfections still reveals the letters in the top coating in (b) to the eye. By subtracting the left- from the right-polarized image (many softwares offer the required subtractive Image Calculator function; for a--c we used Graphic Converter 9 (Lemke software) for d--g we used ImageJ (NIH, USA)) we remove the unpolarized background such that the CSRs appear alone on a black background (c/f). To improve contrast, the color channel corresponding to the reflection wavelength is singled out, the contrast is normalized, and the image mode is switched to monochrome in d/g (done with Graphic Converter 9, using "Effect$\rightarrow$Channels/Frames$\rightarrow$Only Blue in same Window" for c or "Only Red in same Window" for g, followed by "Effect$\rightarrow$Normalize", followed by "Effect$\rightarrow$Black \& White$\rightarrow$Threshold" with the threshold set to 120 for c and 140 for g). Following this principle, a robot equipped with two cameras, imaging simultaneously through right- and left-handed circular polarizers, respectively, could easily detect CSR coatings, even if they are placed in a complex and busy landscape. With two cameras, the two images show the exact same scene (while parallax problems may occur at short distance, they can be corrected for digitally), avoiding problems of moving elements, such as the branch that was failed to be removed in figure~\ref{fig:clc_scales}g due to its minute movement between the photos in d and e.

Interestingly, even under ambient illumination, CSRs appear with largely viewing angle independent color. In fact, the photos in figure~\ref{fig:clc_scales} are not true retroreflection images, \ie there is no collimated illumination of the CSR coatings; they are illuminated solely by ambient light, always providing some light along the viewing direction. The reflection also remains largely circularly polarized, allowing the background subtraction. However, the reflection intensity and thus the contrast will be greater by adding a collimated illuminator with the right wavelength along the viewing direction.

Light illuminating a CSR can also be reflected at other points than the center. Because of the radially varying helix orientation such reflections occur at shorter wavelengths, according to Bragg’s law, and the light is not reflected back to the source but in other directions. If $\theta = 45^\circ$, light from above with $\lambda_r=\lambda_0\cos{45^\circ}$ is reflected into the coating plane, where it hits other CSRs. They reflect the light back to the observer, leading to another highly interesting phenomenon, apparent during near-field observation of a CSR fiducial: an intricate multicolored pattern arises due to this \textit{photonic cross-communication} between CSRs \cite{noh2014tuneable,Fan2015, geng2016high,geng2017elucidating,schwartz2018cholesteric}, see figure~\ref{fig:clc_scales}h. The pattern varies dynamically with changing illumination, the details uniquely defined by the particular arrangement of CSRs. With CSR shells, yet a different type of response, triggered by internal reflection \cite{geng2018through}, is possible (figure~\ref{fig:clc_scales}i). 

As argued in \cite{geng2016high,lenzini2017security}, CSR arrays qualify, thanks to their intricate near-field reflection behavior, as optical \acp{PUF}
\cite{pappu2002physical,McGrath2019}, of great use in object authentication. 
Saying that arrays of CSRs are \acp{PUF} means that their near-field optical response is physically unreproducible, due to the unpredictability of the exact arrangement of the hundreds of individual CSRs in the array. Since the arrangement is randomly defined at production, the probability to produce two exact copies becomes negligible, even if both generate the same far-field fiducial marker pattern, where the individual CSRs are not distinguished. Moreover, since the positions on the scale of individual CSRs are determined by physical process steps that are uncontrollable, it is infeasible to reproduce an exact copy of an existing array of CSRs. Any CSR fiducial is therefore physically unique, and so is its near-field optical response. 
A less beneficial aspect of the blue-shifted reflections at $\theta>0$ is that they also set design limitations for CSR fiducials that should operate in the IR spectrum and need to be unobtrusive to humans \cite{schwartz2018cholesteric,Geng2021encoding}. If $\lambda_0$ is too close to red, we see colors for reflections at $\theta>0$ (figure~\ref{fig:clc_scales}h--i), yet if it is too far into the IR, illumination and imaging requires more complex and more expensive equipment. We will come back to this issue, and how to address it, in section~\ref{subsec:optics}.

The unique combination of fluidity and self-organization into a helical structure in CLCs makes it possible to make CSRs efficiently using fluidic processing, followed by annealing for a time ranging from hours to a few days in order for the helix to align radially throughout (the process can be sped up by osmosis \cite{geng2016high}). While most work has employed a microfluidic processing \cite{uchida2013controlled,geng2016high,myung2019optical} that yields low dispersity (low spread in CSR diameters) but is difficult to scale to mass production, dispersity is not a problem for the applications considered in this paper. In fact, from the point of view of the \ac{PUF} security quality of CSR fiducials, 
dispersity is even highly beneficial. Since it is easy to make CSR droplets with high dispersity in simple stirring procedures \cite{Humar2010} that can be scaled up, and as it was recently demonstrated that CSR shells can also be made from droplets by the use of controlled phase separation \cite{Park2020}, upscaling of CSR production for the use in fiducial markers can thus today be considered a practical problem that is straightforward to solve. To use the CSRs they must finally be turned into a solid, achieved by using reactive molecules such as acrylate or epoxide LC-formers and chiral dopants. By incorporating a photoinitiator, polymerization into durably solid CSRs can be triggered at will by UV light irradiation when the desired optical properties are stable \cite{Asshoff2015, geng2016high,lee2017structural,myung2019optical,Park2020}.

\section{Digital Representation of Environments and Benefits of CSR-Enabled Links to the Physical Counterparts}\label{digrepresent}
The construction of buildings and infrastructure has always been one of the most lucrative industries in the world~\cite{constructionindustry},  
but it has been slow to change, as one of the least digitized industries~\cite{agarwal2016imagining}. Today we start to see a shift towards the deployment of autonomous robots, a development that has been sped up by the covid-19 pandemic \cite{robotconstruction}. The timing for introducing CSR fiducials in \ac{AEC}---including the ability for robots to carry out tasks in construction---is thus ideal. The built environment is not only a function of structural complexity, but of time as well. Projects in \ac{AEC} are impressive when done in under a month, but it is not uncommon that they take 10 years or more. While there are many ways to break down the life cycle of construction, one may succinctly summarize the key stages as: \cite{den-lca,aia-lca}: 

\medskip
\smartdiagramset{back arrow disabled=true, uniform color list=gray!60 for 4 items, text width = 3cm, module x sep = 3.80cm}
\begin{center}
{\smartdiagram[flow diagram:horizontal]
{\textit{Material Manufacturing (Product)}, \textit{Construction Process}, \textit{Use \& Maintenance}, \textit{End-of-life}}}
\end{center}
\medskip

\noindent
These stages (which are not without overlaps) are illustrated with contextual examples in figure~\ref{fig:bim_scales}, also highlighting the role of \ac{BIM} in coordinating the complex and lengthy overall life cycle.

The \textit{ Material Manufacturing (Product)} stage can be subdivided into Concept, Planning, and Design. Coordination, as provided by BIM, plays a central role from the very beginning, facilitating the \textit{Construction Process} in the next stage. The \textit{Use \& Maintenance} stage is concerned with the behavior of occupants (humans and robots) and how they interact with and navigate the space. Naturally, the utilization of the space is informed and guided by the initial \textit{Product} stage. Finally, the \textit{End-of-life} stage is when deconstruction of the environment happens. This stage is similarly informed through the coordination of how the environment was constructed to ensure (as much as possible), that materials can be properly recycled and safely disassembled. Many traditional construction processes for large-scale environments are not designed for easy deconstruction, and are often simply destroyed with explosives. In the following we discuss how \ac{CSR}s may help tracking of physical assets in the various stages, the support for robotic navigation coming naturally into focus in sections \ref{constructionstage} and \ref{inuse}.

\subsection{The Product Stage and the Need for Digital Tracing and Coordination}\label{subsec:product}
Across the manufacturing and construction industries, reliable coordination and tracking of materials, components and operators throughout the supply chain are recognized as critical needs. In the modern \ac{AEC} industry, as well as in deconstruction and recycling, accurate tracking is vital for safety and quality, avoiding health hazards introduced in the past (\eg lead paint, asbestos), ensuring appropriate material quality (\eg structural steel), and that personnel have certified skills and training. On the digital side, blockchain is currently considered as a key technological solution. Multiple methods have been described for integration of blockchain technologies with \ac{BIM}~\cite{mathews2017bim,li2019blockchain} to coordinate the design and construction process of the built environment and maintain a digital representation of the physical reality and the persons acting in it. The application of blockchain is often with respect to a ledger, containing contracts that are integrated with \ac{BIM}. 

A weakness of this process, in its current incarnation, is the lacking ability to track the physical material itself. Coating components with unobtrusive CSR fiducials would provide this missing link. Although the concept of making fiducial markers for AR and robotics not visible to the human eye has been conceived before, implementation details either do not exist~\cite{haas2018unobtrusive} or use primitive means such as IR ink~\cite{lim2016mobile,park2006invisible}. All prior art lacks the unique properties of selective reflectivity and polarization, a vital aspect to the wealth of information that can be communicated by CSR fiducials. Furthermore, previously presented methods lack the multiplicity in function on different scales \cite{schwartz2018cholesteric}. In fact, it is in combination with BIM that the full value of CSR fiducials would come to fruition, embracing---first---the utility in far-field read-out for \textit{categorization} based on the mass-producable fiducial marker pattern generated by all the CSR retroreflection dots connected together (figure~\ref{fig:clc_scales}c/g). Second, near-field read-out of the detailed optical response at the resolution of individual CSRs (figure~\ref{fig:clc_scales}h--i), unique to every individual CSR fiducial, would be utilized for reliable \textit{authentication} \cite{geng2016high,lenzini2017security,schwartz2018cholesteric}. Because one and the same coating, intimately bound to the carrier, combines both functions, CSR fiducials thus provide a highly valuable tool for secure verification of authenticity \textit{and} easy categorization, making them the ideal physical--to--digital link for BIM.

Secure authentication has become a critical component of modern supply chain management, the construction sector being no exception. In fact, the unfortunate reality of the vast financial opportunities in AEC have led to it being one of the most corrupt sectors, with a wide range of issues~\cite{chan2017corruption}. For example, despite efforts put into building code and computational tools to ensure safe and reliable infrastructure, a surprising but real concern is the use of counterfeit or other substandard materials in construction~\cite{engebo2017perceived}. Blockchain is discussed as a way to prevent counterfeit material usage \cite{pishdad2020blockchain}, but the link to the physical objects is missing. CSR fiducials could provide such a link of exceptional value, indeed being unique for each component.

 Although the application in fiducial markers requires us to keep the retroreflection wavelength $\lambda_0$ and thus the helix pitch $p$ reasonably constant among the CSRs in a certain marker, we can randomly mix right- and left-handed CSRs in the marker. This does not affect the analysis of the far-field fiducial marker pattern, which remains as reliable as when only a single handedness is used, as in figure~\ref{fig:clc_scales}, by taking the absolute values of all pixel intensities after subtraction of right- and left-polarized images. We add further unpredictability by using CSRs with varying diameter and (in case of asymmetric CSR shells rather than droplets) thickness. Identification of the unique identity of each CSR fiducial, given to it upon creation, would at the first level amount to locating right- and left-reflecting CSRs at a scale that can be scanned rapidly by optics no more advanced or costly than in a regular mobile phone camera, only adapted for near-field imaging. This scanning could be done at high throughput during manufacturing of the building elements, even as mundane as bricks, and each unique pattern would be recorded in the BIM-coordinated ledger. For higher precision, for instance for more costly elements made out of exceptional material qualities, 
the full behaviour that gives arrays of CSRs the quality of \acp{PUF}
would be observable 
in a microscopic investigation that measures the sizes and orientations of CSRs and classifies the dynamic photonic response  to various illumination conditions. In the language of digital security, multiple challenge--response combinations would be recorded \cite{geng2016high,lenzini2017security,schwartz2018cholesteric}. In this way, every CSR fiducial encapsulates an identifying space of unique optical patterns large enough for satisfying any practical requirements for identification.

The uncloneability, versatility, and deep integration into the object as a strongly adhering coating are the main advantages of CSR fiducials compared to authentication solutions in use today, 
relying on \eg Radio Frequency Identification (RFID) technology \cite{carley2016technologies}, inkjet printed photonic crystals \cite{nam2016inkjet} or paper infused with fibers \cite{chen2005certifying}.
All these currently used technologies have limitations (e.g., they are prone to be cloned, removed, or silenced) that CSR fiducials, thanks to their behavior as \acp{PUF}, show promise in overcoming. We base this conclusion on our arguments presented in \cite{lenzini2017security}, suggesting that CSR fiducials are physically unclonable, their near-field optical response is hardly reproducible in full detail using other materials, and, if realized in form of a coating, they are tamper evident and therefore cannot be moved from one object to another. Besides, the near-field optical response of a CSR array is unique, which makes CSR fiducials or other CSR arrays secure and reliable identifiers, with use in anti-counterfeiting and track-and-tracing applications \cite{geng2016high,schwartz2018cholesteric}.

\subsection{The Construction stage: improving efficiency and worker safety with CSR fiducials}\label{constructionstage}
The use of robotics in \ac{AEC} has been both of research and industrial interest for decades~\cite{kangari1989potential,paulson1985automation}. While early analysis of robotics in construction suggested some use of robotics on-site, a major challenge has continued to be the ability to localize and \textit{understand} the robots' surroundings in a dynamic environment~\cite{feng2015vision}. Moreover, cost made it seem unlikely that, for example, robots would be laying bricks~\cite{richard2005industrialised}. While many of the issues still exist, researchers have continued to work towards on-site robotic construction, \ie employing autonomous robots that are present on the construction site, actively involved in the building process~\cite{schwartz2016use,dorfler2016mobile}. The idea of leveraging \ac{BIM} with information about an environment has similarly been explored, although the robot required a geometric understanding to localize~\cite{lundeen2019autonomous}. By using \ac{BIM} with \ac{CSR} fiducials, it would be possible to facilitate robotic navigation based on a global map (the digital twin) with object-specific geometry (\ac{CSR} fiducial-tagged object). This article focuses on how to realize the link between the digital and physical, which could then be used by control algorithms such as dynamic path planning \cite{10.1145/2522628.2522654}.

\begin{figure}
    \begin{center}
    \includegraphics[page=1, width=0.9\textwidth]{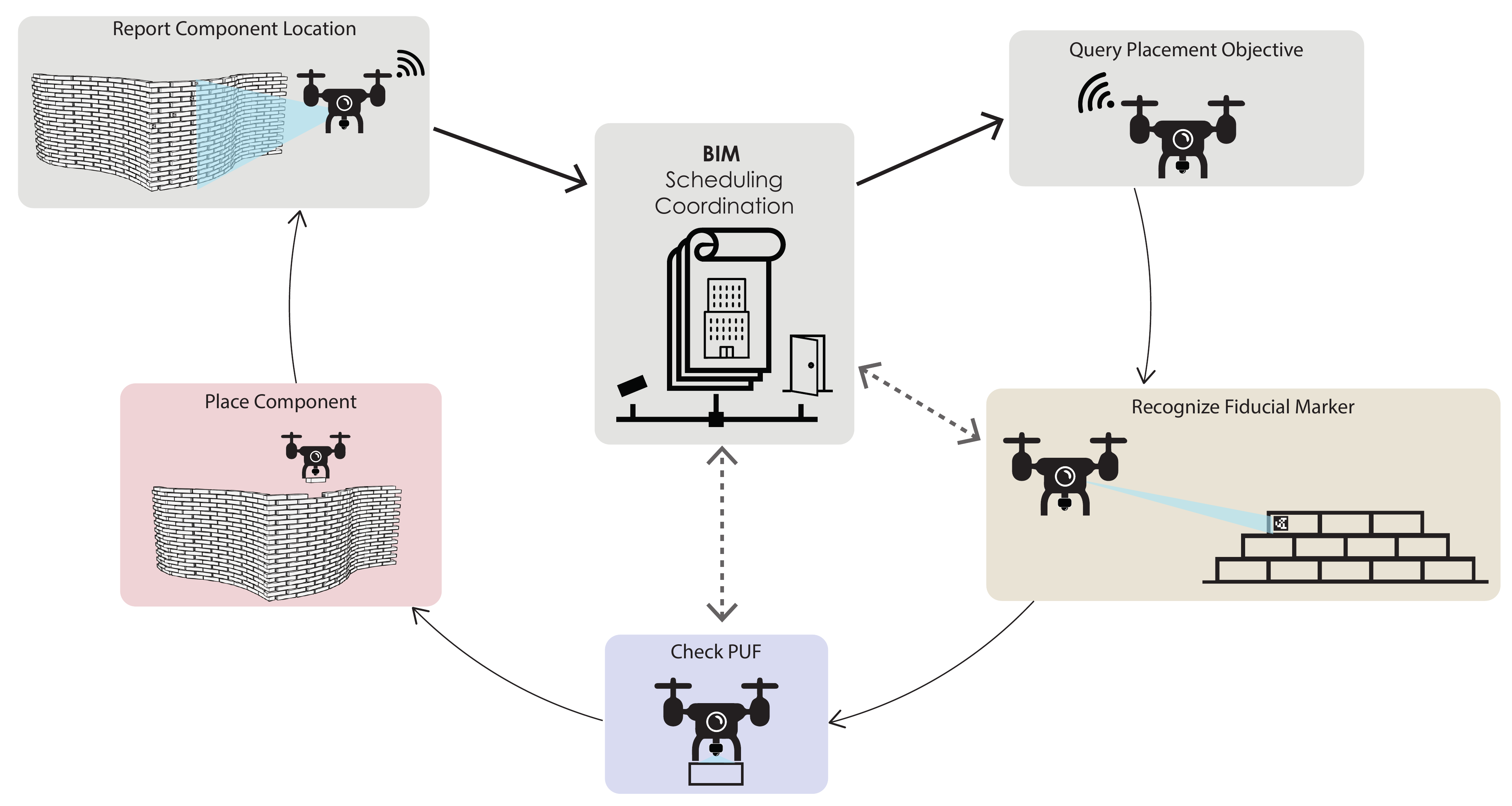} 
     \caption{ Concept of a drone-based construction process. First, the drone queries a scheduling coordinator (\ie \ac{BIM}), to find the next objective. It uses the CSR fiducials on the brick stack to find a brick of the required type. As it picks up the brick, it uses the near-field readout and scans the brick for its unique identity to ensure quality. As it has both the \textit{general object type} (brick of a certain shape, size and material) through the fiducial marker pattern, as well as the \textit{individual brick identity}, the drone is able to communicate back to the \ac{BIM} coordinator which item is placed where, thereby ensuring a proper Digital Twin.}
    \label{fig:droneBrick}
    \end{center}
\end{figure}

With unobtrusive CSR fiducials on all elements on a construction site, from individual bricks to large machines and even workers and robots on site, the vision of robotic construction would become much easier to realize. Inspired by the Ariel Robotic Construction installation using drones to carry foam bricks using optical motion capture~\cite{augugliaro2014flight}, we present a concept in which such a highly automated construction system would be enabled. Thanks to \ac{CSR} fiducials and the information they carry, a robot engaged in construction, as in figure~\ref{fig:droneBrick}, would utilize the far-field retroreflection of CSRs to localize a component it needs for a particular task, for instance a brick, and then approach it for picking it up. Since the component must be recognizable from any direction, multiple \ac{CSR} fiducials should be applied on each component, preferably with more than one on every side. This redundancy improves robustness, as one fiducial may be covered by dust or dirt during construction, but another one on the same surface can be read instead.

As the robot approaches the component to pick it up, a camera and illuminator with near-field imaging capacity attached to its gripper comes into action while the robot is carrying it from the stockpile to the target site. Assuming that a CSR fiducial is placed near the point where the robot grips the item, the camera--illuminator kit can identify the individual brick by probing the spatial distribution of left- and right-handed CSRs and send the result to the BIM server, which compares with the data recorded during the production of the brick (see section \ref{subsec:product}). This also provides the most basic verification of authenticity, applied to every item, because a failed identification would indicate a problem with the origin of the brick. To reach a greater reliability in the verification, a subset of construction elements may have had their CSR fiducials analyzed in full with respect to the \ac{PUF} characteristics, \eg analyzing the photonic cross communication pattern while varying the illumination conditions. The extended time during which the construction robot is in close contact with the element at this stage allows the robot to carry out the full analysis whenever such an item is being moved to its target location.

As the robot also provides the location where each individual item is placed, and in relation to others, the BIM model is complete, having unprecedented data on every individual item in the construction, linking it to its exact producer, material quality and other features. 
The unclonability of CSR fiducials, and the fact that they can become part of the object in the form of a coating, thus building a tamper-evident bind between marker and object, make the system protected against counterfeiting. Should there be any attempt to utilize substandard components in the construction, this will be recorded in the BIM-coordinated ledger, triggering an alert to building inspectors, while also identifying the actor responsible for the error. 

Today's approach to construction does not only suffer from risks of inadequate materials being used in the process, but it also presents hazards for the workers involved. Sadly, construction incurs more occupational fatalities than any other sector, across the world~\cite{constructionsafety}. Some reduction, albeit not elimination of the problems, has been achieved by what is referred to as \textit{prefab}, or pre-fabricated components. In this system, the built environment is approached much like traditional manufacturing in which components are fabricated with high-precision and then transported to a site~\cite{neelamkavil2009automation}. While prefab is one of the most prolific integration of robotics with construction, a deeper integration could have a dramatic positive impact on construction site safety. 

If CSR fiducials were used on a construction (or deconstruction) site, attached on all elements and worn also by workers on their helmets and suits, fatalities due to collisions between workers and machines could be avoided by the improved ability of the machine to detect where humans are and how they move on the construction site. For instance, an automated emergency stop of a crane could be triggered by a CSR fiducial indicating the presence of a person being detected in the crane's path of operation. Personal integrity would be ensured by restricting the use of a single pattern for all workers, to identify them only as humans, not providing any information about who the person is. When appropriate, a few categories of patterns for workers might be used to distinguish between different levels of training and expertise, in order to prevent that workers carry out tasks for which they do not have the necessary skill set. As autonomous robots start appearing on construction sites, such marking-up of all actors and elements may be essential, because the initial development targets robot-\textit{assisted} construction, \ie humans are required to remain on the construction site. This raises a new type of safety concern with respect to the risk of an autonomous robot not correctly detecting a human worker and causing fatal accidents in this way. CSR fiducials on the site would not just make robotic navigation more efficient and accurate, but it would also enable a secondary layer of safety in which robots would easily identify humans on the site.

\subsection{In Use: Enhanced Localization and Information Delivery for Robots and Humans Using AR Devices}\label{inuse}
The marking up of each component, physically and virtually, is valuable not only for the machines involved in construction but, as new built environment emerges, the structure continues to provide data via the CSR fiducials, linking the physical world with precise localization of each component to the virtual database of the \ac{BIM} model. The outward facing fiducials on a completed building facade provide a very large compound graphical code, which can be used by autonomous robots or human-facing Augmented Reality (\ac{AR}) devices (\eg smart phones or devices such as HoloLens) to provide localization. The full potential for augmentation, at multiple scales---from individual products to cities---is represented by the breakout-box of the \textit{In Use} section of figure~\ref{fig:bim_scales}, the most important benefit being facilitated and improved localization for all devices operating in the new environment.

\begin{figure}
    \begin{center}
    \includegraphics[page=5, width=0.8\textwidth]{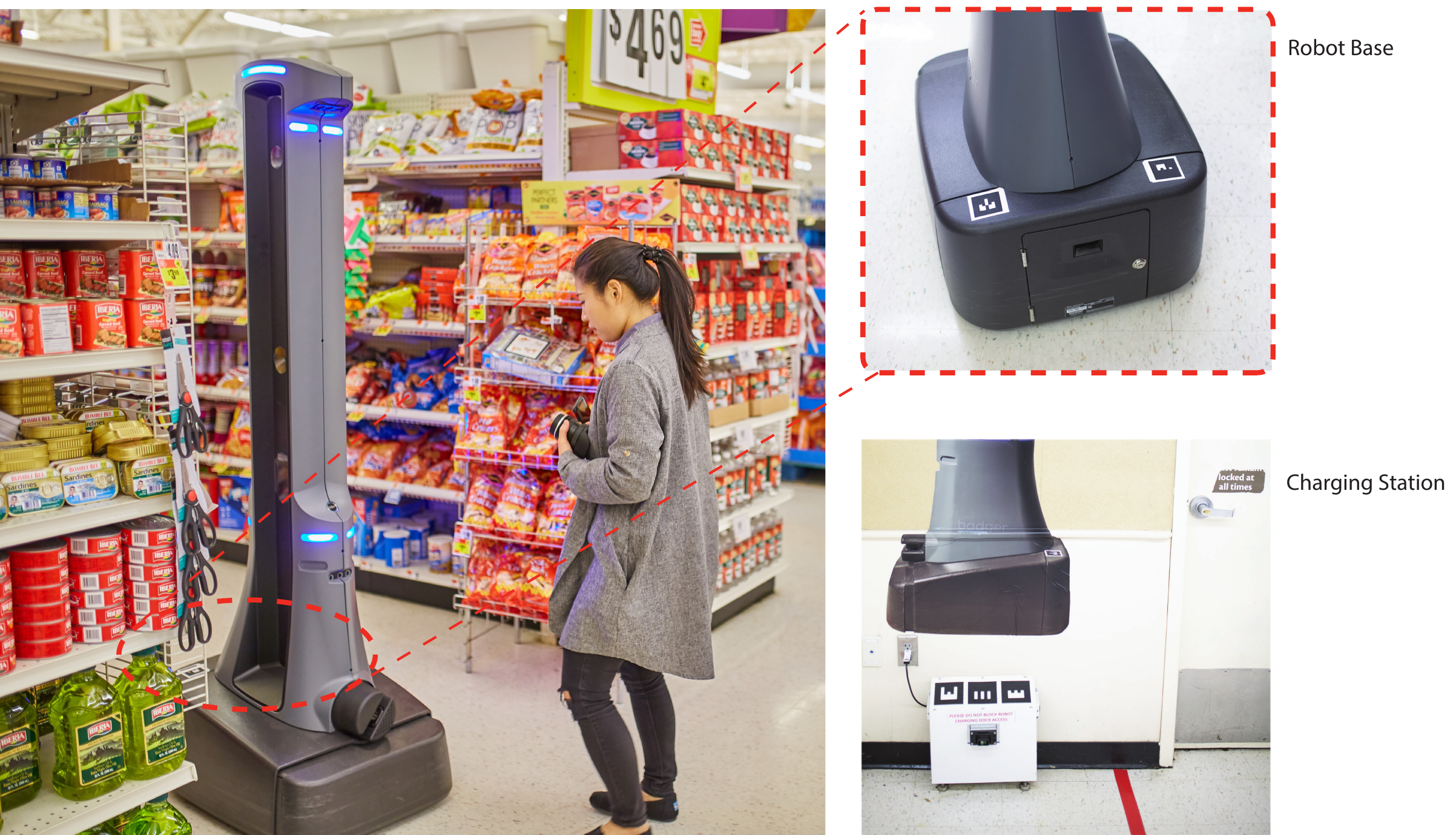} 
     \caption{Robot and customer in a supermarket in New Jersey, October 2019.}
    \label{fig:marketrobot}
    \end{center}
\end{figure}

While localization within a \textit{static} environment may already be considered a straightforward task based on live photography and matching to recorded images related to known locations, this approach may fail if the environment changes or if the environment has not been recorded (\eg pedestrian areas failing to be recognized by Google Maps AR). In an indoor environment, a chair or painting may be part of a recording that defines a specific location in a database, yet the next day these objects may be moved, disrupting the localization algorithm. Similarly, a robot moving around in a supermarket (figure~\ref{fig:marketrobot}) must not be confused because a product had its packaging redesigned. For these and other reasons, object identification and localization by machines is an active research area ~\cite{fuchs2019towards,santra2019comprehensive}, as is the assessment of the reliability of the data~\cite{varadarajan2020benchmark} and evaluation of the computational complexity required to produce it. Fiducial markers of classical (visible) type are today the solution of choice in the very limited number of human-populated environments with sufficiently low expectations on aesthetics to allow them, such as supermarkets. The robot in figure~\ref{fig:marketrobot} has its own fiducial as does its charging station and other key locations. Environments that have been marked up with invisible CSR fiducials as they were constructed could be a game changer in extending this reliable localization solution without limitations. 

Importantly, not only robots but also humans taking advantage of AR devices would benefit. The combination of providing category information as well as localization provided by fiducials has proven integral in the area of \ac{AR} \cite{Garrido-Jurado2014}. Situation-dependent information about a particular door or environment can be linked to CSR fiducials they carry, allowing AR devices to display the information to human users that approach the area. Analyzing the apparent shape and size of each fiducial allows extraction of precise location and orientation with respect to the device, hence the device could continuously update its output such that the information stays in the same position and orientation relative to the user. With invisible CSR fiducials, this solution could be ubiquitously available, of great benefit to future AR implementations. 

Humans are also the direct beneficiaries of improved localization of the class of robots referred to as autonomous vehicles. With the rise of personal mobility vehicles \cite{jeon2015dynamic} that can transfer from outside to inside the built environment \cite{Kim2016d} (\eg offices, hospitals, theme parks, homes), and generalized robots integrating in more aspects of life, the line between where an autonomous car starts, and a robot ends, is continually blurred. For AR as well as for autonomous robotics, the fundamental approach for localization is the same, as both technologies use a computer. Current autonomous vehicle navigation relies on on-board computation, through \ac{AI} and processing of multiple sensor signals \cite{Wei2013b}, matching to \ac{V2V} and \ac{V2I} wireless communication \cite{santa2008} (which includes the opposite direction: infastructure to vehicle). Beyond infrastructure, additional technologies (e.g, vehicle to pedestrian, V2P \cite{v2p}) aim to characterize and identify the surroundings. 

Many state-of-the-art technologies, including the accurate distance mapping of points generated by LIDAR (LIght Detection And Ranging), are problematic in use. 
LIDAR generates sparse dimensional data requiring considerable interpretation through machine learning~\cite{ahmadyan2020objectron}, and while surface penetrating radar \cite{stanley2018method} can localize, it lacks awareness. GPS-based navigation fails in many locations due to lack of signal and it provides no location information in the vertical orientation, causing navigation problems in multi-level structures such as airports or parking garages. The much discussed autonomous vehicle for people \cite{fagnant2015preparing,litman2017autonomous} as well as the more industry established automated guided vehicles (AGV) \cite{schneier2015literature}, suffer from limited understanding of their surroundings. For applications within a building, feasible \ac{V2I} solutions that have been discussed include ceiling codes or magnetic tape, facilitating not only localization but also providing information about the environment \cite{v2i}. Overall, multiple technologies and systems are needed to compensate for flaws \cite{schneier2015literature}. An infrastructure based on unobtrusive CSR fiducials would be a valuable addition, deployed first within public and corporate buildings, then in homes, and eventually in streets and across cities.   

CSR fiducials also provide a reliability and robustness aspect to \ac{V2I} communication, the value of which cannot be overstated. The augmented future of connected devices that motivates this article is frequently envisioned as being enabled by \ac{IOT} devices communicating via the 5G network. Apart from the power consumption issues that this approach would entail if very high resolution is desired, there is also an important national safety issue embedded in this solution. In fact, as we are writing this article, the expansion of the 5G network is being delayed in many parts of the world due precisely to the concerns related to the risk of security breaches in ubiquitous IoT technology and the 5G communication on which it relies. In contrast to IoT devices, CSR fiducials cannot be hacked remotely. We will return to the related issue of resilience to on-site tampering in section \ref{sec:security}.

\subsection{End-of-life}
The critical security and trust issue discussed above, related to the authenticity and appropriateness of materials used in construction, becomes important even at the end-of-life stage. When a structure built using CSR fiducial-labelled elements is torn down, the markers could assist in directing each material to the appropriate waste or recycling stream. The marking-up of each component would thus enable the same coordinated approach of physical and virtual objects being recognized and properly handled, also at this stage of building deconstruction. This would not only have great beneficial impact on the environment, but there can be significant economic consequences, as different waste materials can be connected to value (\eg copper wires) as well as costs (\eg toxic materials). Of commercial as well as societal importance would be the possibility to identify and trace certain high-value or endangered raw materials or products, such as high-performance steel, glass or certain polymers, rainforest wood, or items made out of those materials.

\section{Key Outstanding Challenges in Realizing an Infrastructure of CSR Fiducials}

\subsection{Getting the optics right}\label{subsec:optics}
The peculiar optical properties of CSRs allows them to be detected at the scale of meters, building up a pattern on the scale of cm, and revealing a unique unclonable fingerprint at the scale below mm. This vast range of scales is the enabler for the broad requirements of matching the rich information of a \ac{BIM} dataset. There are however, important research questions to be answered in this space. For instance, the $\theta>0$ reflections involved in the CSR cross communication that is so useful for authentication, also have the consequence that a CSR fiducial designed for IR retroreflection may reflect visible light illuminating the coating from a direction different to that of the viewing direction. This would compromise the requirement of unobtrusiveness of CSR fiducials. Fortunately, the refraction at the air--binder interface limits the maximum angle $\theta$ to values around 65$^\circ$, allowing the design of coatings for IR operation that do not give rise to visible reflections, but the parameter window needs to be carefully evaluated \cite{Geng2021encoding}. With $p$ tuned for near-UV (blacklight) retroreflection, this issue is absent since $\theta>0$ reflections are always blue-shifted, thus they are outside the visible spectrum \cite{Geng2021encoding}.

Apart from their Bragg reflections, CSR fiducials could become visible due to absorption, specular reflection or general scattering. Since properly designed CLCs do not absorb in the visible range, absorption is not a problem, but specular reflection and scattering are of great significance. We can eliminate unwanted scattering as well as internal specular reflection by ensuring that the binder used for the coating is index-matched to the CSRs. Shell CSRs are particularly useful in this respect, as the binder can fill also their interior and thus minimize visibility. Geng et al. recently developed a scalable method to puncture shells \textit{en masse} \cite{Geng2021encoding}, allowing exchange of the original core fluid with the index matching binder, as well as removal of stabilizers from the production process. Near perfect index matching was achieved already with the common UV-curing glue \ac{NOA} as binder \cite{Geng2021encoding}. A combination of experimental characterization and numerical analysis showed that CSR fiducials optimized for IR operation will most likely be weakly detectable under unfavorable lighting conditions due to unavoidable weak scattering, although the visibility can be made low enough for many practical conditions. When using CSRs optimized for UV operation instead, the scattering can be minimized to an extent where the CSRs are practically undetectable by the human eye, provided some further fine tuning of the binder.

\subsection{Printing CSR fiducials}
There are many possible ways of depositing CSRs into the patterns of fiducial markers, and depending on size and context, the method of choice may vary. In some contexts. a gluing of a pre-made film containing a marker may be the best way forward, while in other cases a direct creation of the marker on the carrier object, by patterned deposition of CSRs suspended in liquid binder followed by rapid UV-curing, can be preferable. Since CSRs are quite different from the ink or toner used to print traditional fiducial markers and since the binder must also be deposited in a manner that causes no visible artifacts, there are several interesting engineering challenges in developing reliable and rapid methods to print CSR fiducials, adapted to different use case scenarios. However, we see none of them as fundamental roadblocks, hence we are confident that convenient methods can be developed. 

 \begin{figure}[b!]
  \centering
  \includegraphics[width=\textwidth]{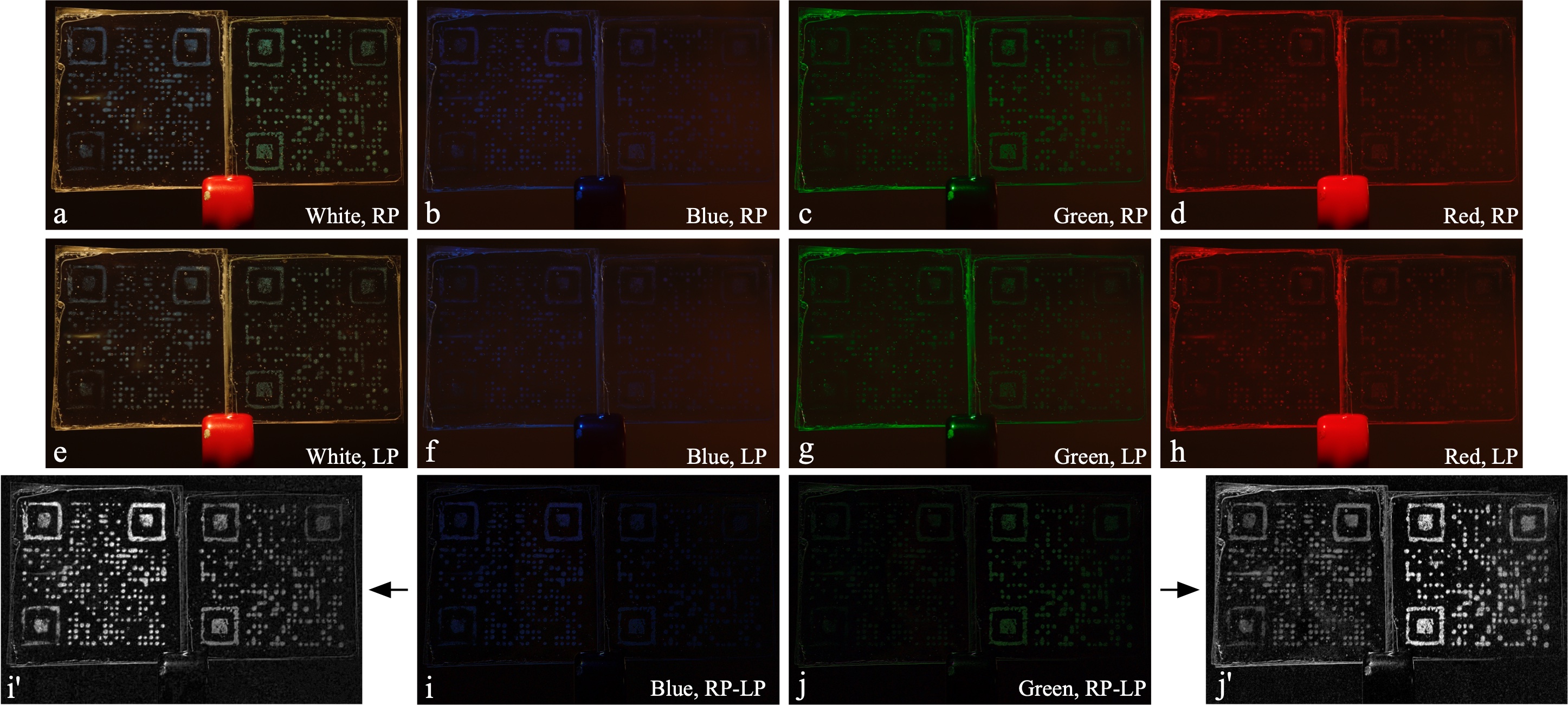}
  \caption{Two QR codes realized in blue- and green-reflecting CSR droplets, respectively, photographed through right-handed (a--d) and left-handed (e--h) circular polarizers. The illumination is white (a/e) and through blue (b/f), green (e/g) and red (d/h) filters, respectively. Under red illumination, all CSRs are almost invisible since $\lambda_0$ in both QR-codes is shorter than the wavelength band of the red filter. Panels (i) and (j) show the result of subtracting the left-hand from the right-hand-polarized images under blue and green illumination, respectively, and (i') and (j') show the corresponding monochrome and contrast-enhanced images. Each QR-code is 6~cm by 6~cm.}
  \label{fig:QR}
 \end{figure}

A first simple demonstration of machine-readable CSR coatings is given in figure~\ref{fig:QR}. A mold in polymethylmethacrylate (PMMA) was created by etching depressions in all areas where CSRs should be deposited. A layer of Parafilm was then worked into the mold until it adopted the pattern of depressions. Dry CSR droplets were sprinkled over the Parafilm--mold sandwich, which was gently shook in order that the CSRs localized into the depressions. The entire area was then covered with \ac{NOA} glue with a refractive index near that of the CSRs, and by UV-curing the \ac{NOA} the coating was completed. The Parafilm could easily be peeled off, allowing the mold to be reused for further marker production.

The resulting coatings, one blue- and one green-reflecting, are far from perfect and they contain many artifacts giving rise to visual noise. Nevertheless, they demonstrate that the comparatively advanced patterns of QR-codes can be realized reliably with CSRs, using very simple methods. And even with visible colors as close as green and blue, the contrast being further reduced by the scattering caused by the use of CSR droplets with imperfect radial CLC helix alignment, sufficient separation by using different illumination filters can be achieved. By turning the blue and green channels, respectively, into greyscale and enhancing the contrast (we applied, equally to the entire image, the effect "Auto levels" of Graphic Converter 9), images of the encoded patterns good enough for regular QR-code readers to detect them are obtained (figure~\ref{fig:QR}i'/j'). Because the definition of the three critical corner squares is not perfect (in part due to incomplete filling with CSRs, in part due to slightly uneven illumination), QR-code readers may have some difficulties detecting the codes immediately. Nevertheless, we have confirmed that a regular smart phone in photo mode directed to panels (i') and (j') is able to detect the codes. The reader may test it, finding that the two codes link to the websites of the first and last author of our paper, respectively. The incomplete separation between blue and green codes can sometimes make a reader detect the 'wrong' code, but the example clearly demonstrates that the potential to define QR-codes---and therefore also fiducial markers---using CSRs is there.

In future use cases, the reflection wavelengths will be in the UV and IR ranges, respectively, rather than in the ranges of different visible colors, which will make the separation much more efficient, not least when employing UV and IR illuminators, respectively. Moreover, by optimizing and automating the CSR fiducial printing process, many artifacts from dust particles etc. can be avoided. Finally, the use of CSR shells rather than droplets ensures perfect radial helix alignment and minimized reflections beyond $\lambda_0$ \cite{Geng2021encoding}, hence we expect significantly better contrast between patterns designed with different $\lambda_0$ than what is demonstrated in figure~\ref{fig:QR}.

\subsection{Lifetime of CSR fiducials deployed outdoors}
While marked-up items and environments in many of the use cases conceptualized in figure~\ref{fig:bim_scales}  belong to the interior, the 10--100~m scale use cases, like outdoor navigation and during construction, will subject CSR fiducials to natural weather. It is unavoidable that extreme weather conditions may temporarily hinder optical communication, but regular wind, humidity, ice, sand, dust and varying temperatures should at least not be detrimental to the long term use. CSRs are fabricated based on mesogens, \ie liquid crystal-forming compounds. Very similar compounds to those used for CSR production are the basis for the highly mature liquid crystal display (LCD) technology, where they are used \eg in TVs, computer monitors, mobile devices and projectors. In this context, liquid crystalline materials have proven their long-term durability under ambient conditions. In fact, even under rather harsh conditions, such as the elevated temperature and continuous intense light exposure that liquid crystals must endure in a projector, these materials show excellent lifetime.

In CSRs, the liquid crystalline state is used only during production, since they are polymerized into solids before being incorporated in the coatings making up the CSR fiducials. This can be expected to further enhance the stability and robustness, as the molecular organization is fixed, showing no temperature dependence (up to decomposition, which for the materials we are considering here is at least up to about 200$^\circ$C), and reduced sensitivity to humidity and light. Nevertheless, for outdoor applications, light stability can be further improved by the addition of photo-stabilizers \cite{Kucuk2018}, such as benzotriazoles \cite{Kuila1999}, which are basically colorless inert dyes that do not disturb the liquid crystal order during the production phase. 

Since the CSRs are incorporated in a binder when used in a CSR fiducial, we also have an excellent toolbox to ensure durability by tailoring the binder, and we may also add an additional top coating if desired. While the binder must first of all provide index matching and a durable solid state during normal conditions, requirements that can be fulfilled by commercially available binders such as \ac{NOA}, we may also consider the use of special protective coatings. These can be designed to absorb the high energy irradiation coming from sunlight \cite{Zayat2007} or artificial light sources. If desired, other functions can also be added, such as scratch resistance \cite{Hwang2003} and self-cleaning \cite{Sethi2018,Feng2018Oscillating}. Such functions may be particularly valuable for application in the harsh environments of construction sites.

\subsection{Reliability and security issues with CSR fiducials.}
\label{sec:security}

From a reliability and security point of view, the first question is if the message carried by a CSR fiducial is equally (or perhaps even more) detectable and readable as when the classical black-and-white fiducial markers are used. Although we have not yet performed a systematic study of CSR fiducial detectability and readability over large distances for mobile robots (we are currently in the start-up phase of such a study), the results in section \ref{csr:sec} provide proof-of-concept evidence on how detection and read-out should work. The demonstrated removal of the background even gives a strong argument for easier and more reliable detection in visually rich environments. Regarding clarity, which requires---in addition to detecting and reading---also flawlessly interpreting the patterns encoded in CSR fiducials despite minor errors, we have no reason to believe that CSR fiducials would bring up different challenges than traditional black-on-white markers, especially if we rely on common marker designs, e.g., Aruco, ARTags, ARTool or VisualCode libraries (\eg see \cite{Mantha:isarc19}).  

Moreover, the ability to locate the signal to the UV as well as the IR spectral ranges may be beneficial in reducing loss of detectability in certain weather conditions compared to traditional fiducial markers, but there are obviously still situations where a device would no longer be able to correctly detect and read out CSR fiducials. For this reason, outdoor autonomous vehicles would always have other navigation technologies working in tandem with the CSR fiducial read-out. And, just as for human drivers, there are weather conditions where the vehicle simply must stop moving until the situation improves. A simpler type of obstruction that affects all fiducials, including those realized using CSRs, is when a mobile object blocks the sight to a fiducial. Here, CSR fiducials have a significant advantage through their unobtrusiveness, which allows them to be deployed ubiquitously, with significant redundancy. A vehicle approaching a building can thus establish which building it is even if a truck is parked in front of it, by reading off the redundant markers, e.g., deployed higher up on the wall. 

It is thus very likely that CSR fiducials live up to all requirements on reliability arising from the way fiducial markers are used today in research or in restricted environments. However, since the step to out-of-lab operation in diverse human-populated environments was never an option with traditional fiducial markers, the much more severe requirements on reliability and security that such usage entails has not yet been studied. What we are dealing with here is the information security, \ie the resilience of CSR fiducials against attempts of tampering, e.g., by players interested in spreading misinformation, disrupting robots' navigation and understanding of their surroundings, and breaking physical--to--digital twinning. At the time of writing~\footnote{\today}, we have not found any article describing research on such fiducial marker manipulation (for QR-codes the situation is very different). Fiducial markers deployed in everyday environments are likely to suffer attacks similar to those targeting QR-codes \cite{Yong2019}. The situation is worsened by the facts that fiducial markers have a limited dictionary space and that there is no place in the marker pattern for error correction codes, integrity check-sums, or other security guarantees that are typically used in the more information-capable QR-codes\cite{Wahsheh2020}. Considering only the marker pattern, it should thus be rather easy to change one message into another without discovery. 

These considerations lead us to believe that an infrastructure with CSR fiducials will adapt the \textit{concept} of fiducial markers but not necessarily the existing marker libraries. To ensure sufficient security and reliability for such an infrastructure to guide the navigation of robots co-inhabiting our everyday spaces with humans, features that allow detection of manipulation of existing fiducials, introduction of forged fiducials, or any other attacks on the infrastructure, should be added. These may be inherent to each individual fiducial, as in QR-codes, or they may utilize some integrity/coherence check with other fiducials that must be in the vicinity---of course in case of an intact infrastructure---as the use cases we are discussing by necessity mean that a fiducial is never used in isolation. The latter approach would be particularly valuable in revealing forged fiducials introduced in an environment, \eg with the purpose to guide autonomous vehicles or other robots down an incorrect path. 

Thankfully, the optics of CSR fiducials offer additional modes for realizing security and reliability that go beyond the pattern design. One could, for instance, overlay different patterns in a single fiducial, each pattern realized using CSRs with different polarization, or $\lambda_0$, or both. This could be utilized, for instance, to depict the main fiducial marker pattern in the IR spectrum while a related security pattern (linked to the fiducial itself or to its surrounding fiducials) becomes visible only under UV illumination. 

Even without advanced approaches like overlapping patterns, CSR fiducials are superior to traditional fiducial markers in terms of information security, thanks to their fundamental optical properties. In a traditional printed fiducial marker, as well as in a QR-code, white pixels can be blackened and black ones can be whitened. If digitally generated using a display, codes can be modified by applying a mask. The effect will be to have the message changed according to an adversary's intentions, a strategy used in phishing attacks \cite{Yong2019}. QR-codes suffer also from code-in-code attacks, consisting in adding a smaller QR code in a non-coding zone within a larger code. 

Codes can be also covered entirely with new codes, thus replacing the original message by a corrupted one. Being realized in the UV or IR spectra with circularly polarized light, the pattern of a CSR fiducial is much more challenging to modify. The basic action of applying a mask would imply being able to change the polarization or wavelength of the reflection of existing CSRs and/or create a CSR-like response from regions where no CSRs are placed, all without compromising the omnidirectional retroreflectivity of the overall fiducial. To anyone with experience in producing CSR fiducials, this task appears as practically impossible. 
The fact that the CSR fiducial is undetectable to the eye will not help the adversary, who needs dedicated equipment to visualize the authentic fiducial as well as the modifications made. Incomplete knowledge about which wavelengths and polarizations are used in the fiducial, \eg in case of multiple overlaid codes, would lead to failure in the manipulation attempt.

A more likely mode of tampering would be that the adversary gains access to the equipment of making CSR fiducials, introducing fake markers in the environment and possibly blocking off existing ones with a coating that is opaque to wavelengths around $\lambda_0$ of the used CSRs. With such equipment, the attack could be deceitful enough to compromise the navigation of a robot or AR device entering the manipulated environment, if only traditional fiducial marker patterns are used and if only the far-field response is analyzed. To allow detection of the manipulation by far-field read-out only, an enhanced marker design should be employed, as suggested above, although cross correlation with the BIM server would also detect the problem, albeit requiring radio communication or a pre-downloaded set of the BIM database. If near-field inspection is possible, for instance by a dedicated robot at regular intervals, this type of attack, involving the silencing of authentic fiducials and introduction of faked ones, would be revealed, as the unclonability of CSR fiducials prevents fake fiducials from replacing the original ones \cite{lenzini2017security}. A full rigorous analysis of the security of CSR fiducials in various scenarios is planned in our future work.

\acresetall 
\section{Discussion and Outlook}

The promise of \ac{AR} and autonomous robots has enormous societal meaning and strong scientific backing, yet at even the simplest of forms has been greatly underwhelming in implementation. While research for years has shown localization through image processing and concepts of \ac{AR}-based experiences in the built environment, key challenges are yet to be addressed at a resolution affording mass adoption and suitable experience. From our interdisciplinary perspective, a key issue has been discipline-limited solutions to large scale problems. As engineers and computer scientists focus on methods and algorithms for understanding the environment, the task of \textit{changing} the environment has remained with architects and designers largely unaware of the engineering challenges---and opportunities---necessary for a connected environment. In the past couple of years, the rapid increase of industrial and academic interest in connecting the physical and virtual (\ie Digital Twinning) shows promise of a multi-disciplined solution. We propose the use of \acp{CSR} as a critical multifunctional material for tying together the multiple, and converging, ideas around \ac{BIM}, \ac{IFC}, \ac{PLM}, and \ac{AR}. As we have argued above, CSR fiducials can be used to reliably link physical elements to their Digital Twins, enhancing security of the built environment, improving safety during construction, and affording high-quality localization during use and effective end-of-life recycling and disposal of building components when needed.

In our past work~\cite{schwartz2018cholesteric} we laid the foundation for a visionary concept that described unique properties of \ac{CSR}s, and how they can facilitate an information-rich environment. Now, we have demonstrated coatings with CSRs laid out in patterns more complex than those of fiducial markers, encoding information that can be quickly revealed and highlighted over a distance of several meters by a simple and efficient background subtraction. We also initiated a discussion of the security and reliability features of CSR fiducials, highlighting their advantages also from this point of view compared to regular fiducial markers. More research is still needed regarding the manufacturing of CSRs and of CSR fiducials in order to reach the required mass production yields. Importantly, CSR size dispersity is a feature, not a bug, in the contexts we consider, hence the currently widely used microfluidic production method can be replaced with high-yield emulsification methods like stirring or shaking. The recent demonstration of CSR shells developing from droplets following controlled phase separation \cite{Park2020} is important in this context, as CSR shells are the best choice to ensure rapid annealing of the radial helix structure \cite{geng2016high} and minimum visibility of the final CSR fiducials \cite{Geng2021encoding}. 

The most critical challenge from a materials development point of view will be to maximize functionality while minimizing detectability by the human eye. Even though visible Bragg reflections can be avoided by careful tuning of the helix pitch and embedding in appropriate binder \cite{Geng2021encoding}, non-selective scattering may be impossible to prevent entirely, due to the anisotropy of the CSRs. With an optimally index-matched binder, UV-reflecting CSR fiducials can be made practically undetectable to the human eye, but IR-reflecting fiducials may remain slightly visible on certain backgrounds under particular lighting conditions \cite{Geng2021encoding}, and in some contexts this will not be acceptable from an aesthetic point of view. Other important material-related aspects are costs and adherence to the intended carrier of CSR fiducials. While it would greatly facilitate construction if every brick could be marked up, as envisioned in figure~\ref{fig:droneBrick}, a porous brick surface may render reliable attachment of the fiducial very challenging. Moreover, the cost of the CSR fiducial must be truly negligible to motivate marking up such common elements, hence there may be a minimum value of elements for motivating the application of a fiducial.

\subsection{Expanded View and Future Possibilities}

With the origins of Digital Twins in \ac{PLM}, an expanded idea of twinning not only the components directly linked to \ac{AEC} and building life-cycle, but also to the objects within, is not unwarranted. Traditional \ac{PLM} could find valuable use of \ac{CSR} fiducials, assisting recycling of valuable components and safe disposal of harmful materials when a product has served its purpose, thereby supporting the circular economy. In addition to the value of twinning a component, the \ac{CSR} fiducial could be specifically designed to exhibit a functional change when subjected to unwanted conditions. For instance, the binder in which the CSRs are integrated could be chemically designed such that it looses its shape fidelity above a certain temperature, in order to reveal component failure due to, or triggering, overheating. If a non-crosslinked polymer forming a glass at the normal usage temperature is used as binder, it would melt into a viscous liquid if the tag is heated above the glass transition temperature $T_g$, contracting towards a droplet shape and thus losing the locked-in arrangement of the CSRs. The permanent loss of CSR fiducial function would then allow easy identification of the faulty component. Since we can also design the binder for very low $T_g$, the same concept could be used to detect breaches in a cold chain, currently gaining widescale attention due to the importance in shipping of vaccines. The binder can also be designed to resist only certain amounts of mechanical strain, by using varying degree of crosslinking, such that mechanical failure can be easily detected. By applying the CSR fiducial on particularly failure-prone regions of a construction element, this may help the identification---and consequent rejection---of faulty elements also during construction.

If very high-yield and low-cost production of CSR fiducials can be realized, one may even anticipate marking up individual products on store shelves. Markers undetectable by the human eye could cover the full packaging, making it easy for suitable devices to detect and interpret them regardless of orientation. This would present a significant advantage to today's printed bar or QR codes, which require a user to pick up the item and place it at the right distance and orientation for a device to read it. Robots could thus easily monitor the stock in the shelves, and a customer holding an \ac{AR} device could be alerted if a product contains substances to which they are allergic or if it does not live up to the person's expectations related to product origin. All while the device provides general help in finding items the customer is looking for. Such a perspective is certainly far into the future, and its realization would require many practical challenges to be overcome.

Accurate and secure identification of products offers new modes of accessibility and user experience in daily life. As with robots, humans are often hindered by the built environment, regarding physical accessibility (hence the importance of autonomous mobility) or visual and mental understandings of the space they occupy. In essence, a \textit{screen reader}---for the world---enabled by BIM and CSR fiducials would be created through a mixture of AR devices such as a phone. The integration of such a technology could help the visually impaired not just in their own navigation, but in product identification from both a store shelf and in their own home. In a universal application, the assistance in assembling something (e.g., IKEA furniture), could be augmented not just by simple graphical instructions, but by device-identifiable product components, aesthetically invisible to the human. 

Finally, the use of well-deployed CSR fiducials can have life-saving applications outside the connection to a virtual asset. Similar to the usability of \ac{CSR} fiducials for construction worker safety, they could be implemented in jackets or life vests (like in HV clothing) or in an external coating of an airplane or a ship. These codes could then be used to locate persons or vessels from the air in emergency situations, like when survivors must be found after a plane crash or when a hijacked merchant vessel is searched for from the air. Thanks to the retroreflectivity and ability to remove the unpolarized reflections from the sea, the search could take place day as well as night, with much increased likelihood of finding the objects or persons.

\medskip

\medskip
\textbf{Acknowledgements} \par 
We thank Dr.~Jose Luis Sanchez-Lopez and Prof.~Holger Voos for valuable discussions concerning the implementation of CSR fiducials in robotic navigation. We gratefully acknowledge financial support from the European Research Council (ERC, Proof of Concept project VALIDATE, grant code 862315), the Office of Naval Research Global (project LAB'RINTH), and the Luxembourg National Research Fund (C17/MS/11688643/SSh).

\medskip

\bibliographystyle{iopart-num-long}
\bibliography{references}
\end{document}